\documentclass{article}
\usepackage{spconf,amsmath,amssymb,graphicx,hyperref}
\usepackage{booktabs}
\usepackage{multirow}
\usepackage{makecell}
\usepackage{float}

\usepackage{afterpage}

\title{LITECASS: A LIGHTWEIGHT END-TO-END NETWORK FOR REAL-TIME STEREO CINEMATIC AUDIO SOURCE SEPARATION}

\name{Yuanxin Guo$^{1}$, Qiang Ji$^{2}$, Mengmei Liu$^{2}$, Yuhan Lv$^{1}$, Ningning Pan$^{1\star}$\thanks{Corresponding author: nnpan@swufe.edu.cn}, Gongping Huang$^{3}$}
\address{$^{1}$Southwestern University of Finance and Economics, Chengdu, China \\
         $^{2}$Xiaomi Automobile Co., Ltd., Beijing, China \\
         $^{3}$Wuhan University, Wuhan, China}

\begin{document}
\maketitle

\begin{abstract}
Cinematic audio source separation (CASS) decomposes a soundtrack into dialogue, music, and sound-effects (SFX) stems. 
Existing CASS methods, however, suffer from two critical limitations: they rely on heavily parameterized network architectures and GPU-class hardware, limiting their use in real-time and resource-constrained scenarios, and they are overwhelmingly designed for monaural signals, leaving the stereo scenario largely unexplored.
We present LiteCASS, to our knowledge the first lightweight end-to-end network for real-time stereo CASS. 
LiteCASS combines deterministic STFT subband rearrangement with two jointly trained compact U-Nets: the first extracts dialogue, and the second separates music and SFX from the predicted non-speech component. 
A multi-task waveform-domain $\ell_1$ loss supervises all stems. On a spatialized stereo extension of DnR\,v3, LiteCASS-K8 uses only $1.06$\,M parameters and $0.72$\,G MACs per second, achieves the highest averaged SI-SDR among the compared CASS baselines.
\end{abstract}


\section{Introduction}

\begin{figure}[t]
    \centering
    \includegraphics[width=0.5\textwidth]{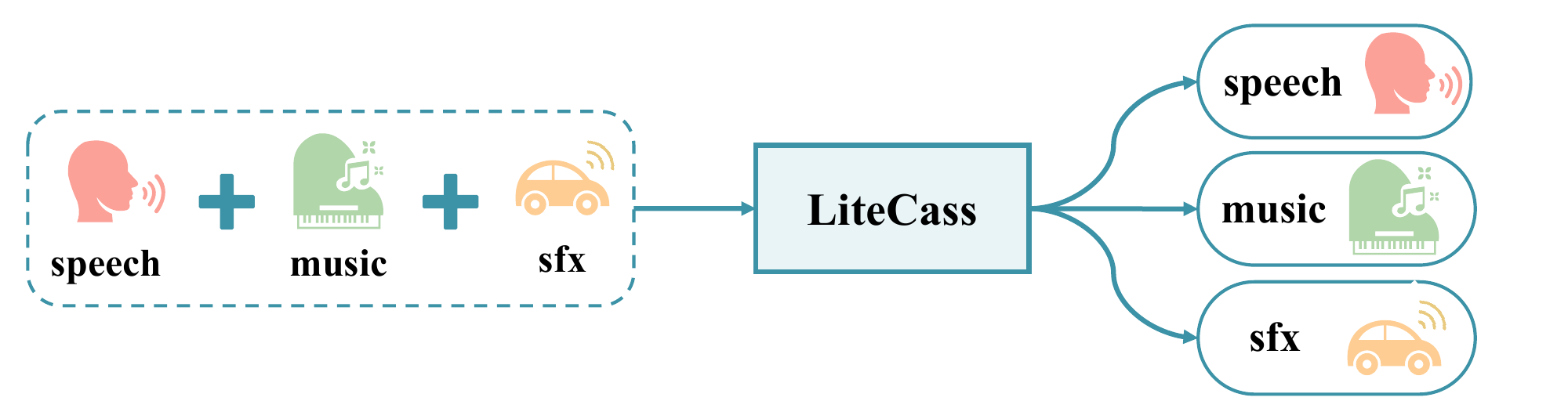}
    \caption{
        Illustration of the Cinematic Audio Source Separation (CASS) task.
    }
    \label{fig:cass_task}
\end{figure}

Cinematic audio source separation (CASS) aims to decompose a film soundtrack into dialogue, music, and sound-effects (SFX) stems~\cite{casstask, tacklingcocktail}, as illustrated in Fig.~\ref{fig:cass_task}. 
Compared with speech or music separation, CASS is more heterogeneous, since dialogue is usually sparse and spectrally structured, music spans diverse instruments and production styles, and SFX forms a long-tail class that includes nearly all non-speech and non-music events. 
This heterogeneity makes CASS both useful on dialogue enhancement~\cite{dialogueenhancement}, subtitle generation~\cite{subtitle}, multilingual re-dubbing, automatic post-production, content-aware loudness control~\cite{loudnessnormalisation}, and assistive listening on consumer devices, and difficult to address under tight computational budgets.

\begin{figure*}[!t]
    \centering
    \includegraphics[width=\linewidth]{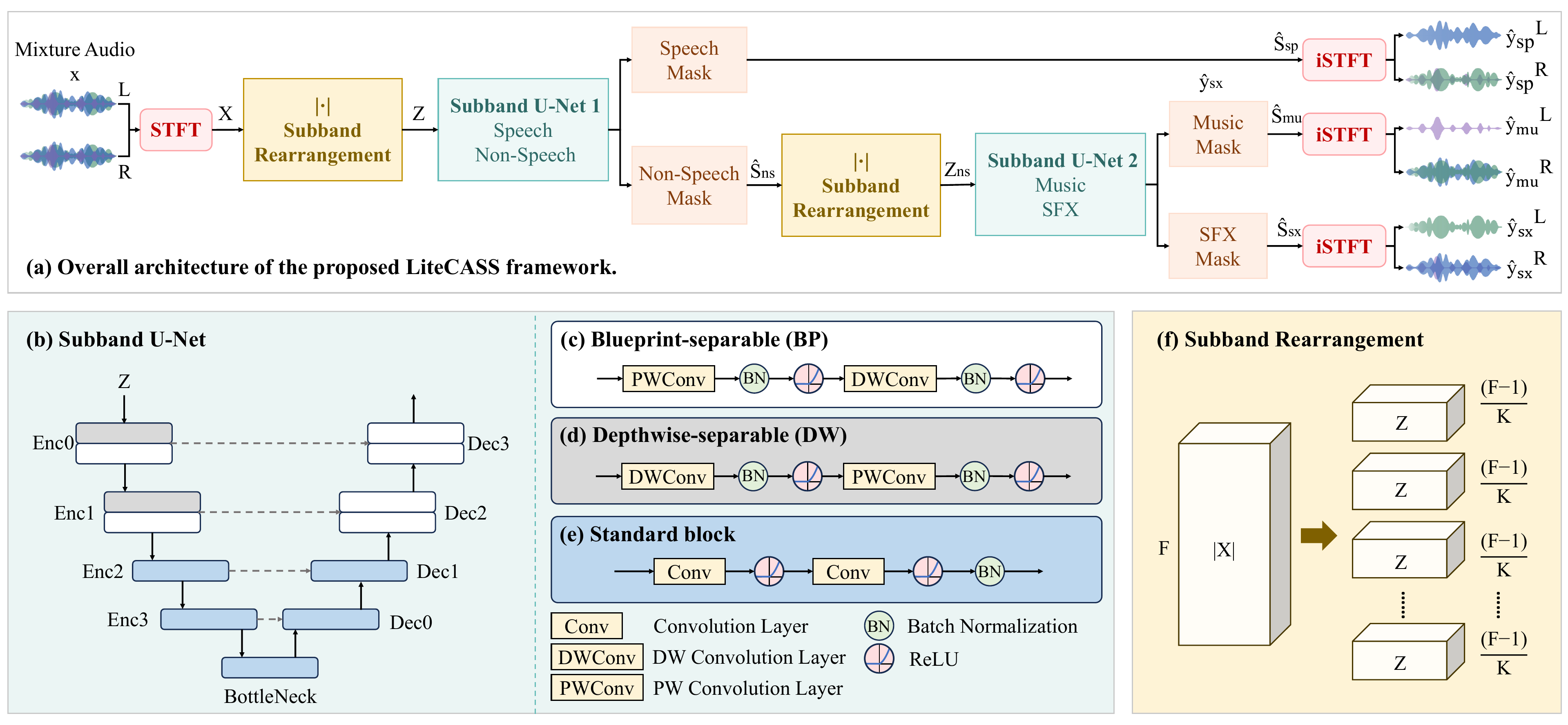}
    \caption{Detailed architecture of LiteCASS:
    (a) Overall framework, (b) Subband U-Net, (c) Blueprint-separable (BP) block,
    (d) Depthwise-separable (DW) block, (e) Standard block,
    (f) Subband Rearrangement.}
    \label{fig:litecass_network}
\end{figure*}

Recent neural systems have substantially improved CASS. MRX~\cite{mrx} formalizes CASS as the \emph{cocktail fork problem} and fuses multiple short-time Fourier transform (STFT) resolutions~\cite{crossnetunmix} around a bidirectional LSTM stack with a per-source, per-resolution multi-decoder. Bandit~\cite{bandit} adopts a common-encoder, three-decoder band-split design~\cite{bsrnn} that models time and band axes with residual GRUs~\cite{gru}. AV-CASS~\cite{avcass} casts CASS as audio--visual conditional flow matching, generating the three stems under a fused visual condition. General waveform separators such as Demucs and its hybrid and transformer variants~\cite{demucs, hdemucs, htdemucs} are also widely used. These systems achieve high accuracy, yet they suffer from two critical shortcomings that have not been jointly addressed.

First, their recurrent, transformer, or large convolutional modules incur high parameter counts and computation, typically requiring GPU-class hardware for real-time inference, which are undesirable for live broadcast, interactive editing, mobile deployment, and embedded playback. 
Although lightweight separators~\cite{convtasnet, tcn, ultralitess, fspen, ulunas} do exist, they are designed specifically for speech or music and do not handle the heterogeneous dialogue–music–SFX structure of cinematic audio. 
Second, and equally important, existing CASS research is predominantly conducted in mono~\cite{dnrv3}, whereas cinematic soundtracks are commonly delivered in stereo or multichannel formats. 
A truly practical stereo CASS system must not only separate the three sources but also preserve their spatial cues, a requirement that remains largely unexplored in the literature.

To fill this gap, we propose LiteCASS, a lightweight end-to-end network for real-time stereo CASS, with three contributions: 1) to the best of our knowledge, LiteCASS is the first end-to-end system explicitly designed and evaluated for stereo CASS, using a reproducible spatialized version of DnR\,v3; 2) a compact subband U-Net that folds STFT frequency bins into channels through parameter-free subband rearrangement and reduces computation with separable convolutions. 

\section{Method}
\label{sec:method}

\subsection{Problem formulation}
Given a stereo mixture waveform $\mathbf{x} \in \mathbb{R}^{C \times T}$ with $C=2$, LiteCASS estimates the stereo speech, music, and SFX waveforms, denoted by $\hat{\mathbf{y}}_{\mathrm{sp}}$, $\hat{\mathbf{y}}_{\mathrm{mu}}$, and $\hat{\mathbf{y}}_{\mathrm{sx}}$, respectively.
LiteCASS performs separation in the STFT domain, using the complex spectrogram $\mathbf{X} = \mathrm{STFT}(\mathbf{x}) \in \mathbb{C}^{C \times F \times L}$, where $F$ and $L$ denote the numbers of frequency bins and time frames. The network uses magnitude spectrograms as input features and predicts real-valued masks. These masks are applied to complex spectrograms, and the final source waveforms are reconstructed by inverse STFT.

\subsection{Overall structure of LiteCASS}
As shown in Fig.~\ref{fig:litecass_network}~(a), LiteCASS uses two jointly trained Subband U-Nets in an end-to-end hierarchy: $h_{\theta_1}$ maps $\mathbf{X}$ to $(\hat{\mathbf{S}}_{\mathrm{sp}}, \hat{\mathbf{S}}_{\mathrm{ns}})$, and $h_{\theta_2}$ maps $\hat{\mathbf{S}}_{\mathrm{ns}}$ to $(\hat{\mathbf{S}}_{\mathrm{mu}}, \hat{\mathbf{S}}_{\mathrm{sx}})$. Music and SFX losses therefore also shape the upstream speech--non-speech decomposition.

This hierarchy matches the structure of cinematic audio: speech is first separated as a sparse and spectrally structured component, while music and SFX are separated within the predicted residual non-speech component. 

\subsection{Subband U-Net separator}
Both decomposition levels use the same compact Subband U-Net separator (Fig.~\ref{fig:litecass_network}~(b)) as the mask estimator, which is designed to reduce the cost of two-dimensional convolution on full-resolution STFT magnitudes while preserving frequency-dependent modeling. Given a complex spectrogram, the network takes its magnitude as input features and predicts real-valued masks that are later applied to the corresponding complex spectrogram.

A direct U-Net over the full STFT magnitude is computationally expensive because the frequency axis is large. LiteCASS therefore applies a deterministic subband rearrangement before convolutional processing, as illustrated in Fig.~\ref{fig:litecass_network}~(f). After removing the Nyquist bin, the remaining frequency bins are divided into $K$ contiguous groups of equal size and folded into the channel dimension:
\begin{equation}
\label{eq:subband}
|\mathbf{X}| \in \mathbb{R}^{C \times F \times L}
\;\xrightarrow{\;\text{subband}\;}\;
\mathbf{Z} \in \mathbb{R}^{KC \times F' \times L},
\end{equation}
where $K$ denotes the number of subbands and the reduced frequency dimension is $F'=(F-1)/K$. This operation discards no spectral information and has no trainable parameters; it merely converts a high-frequency-resolution representation into a compact subband-channel one. As a result, convolutional layers operate on a frequency axis that is $K$ times shorter, while the folded channels let the network jointly model different subbands and stereo channels. Before mask application, the inverse rearrangement restores the original frequency layout, and the removed Nyquist bin is recovered by padding.

The backbone is a compact encoder--decoder U-Net with skip connections. The encoder takes the $KC$-channel subband representation as input and consists of five convolutional blocks, each containing two convolutional layers that keep the time--frequency resolution unchanged. Downsampling is performed by $2\times2$ average pooling after each block except the deepest one, which retains its resolution; the feature dimension is correspondingly increased along the encoder. The decoder mirrors this process with $2\times2$ transposed convolutions and skip connections from the corresponding encoder blocks, which help preserve high-resolution time--frequency details for mask estimation.

To reduce computation in the most expensive high-resolution layers, LiteCASS uses separable convolutions in the two outermost blocks of both the encoder and the decoder, while standard convolutions (Fig.~\ref{fig:litecass_network}~(e)) are retained in the inner blocks, where feature maps are smaller and the computational cost is less dominant. Within these separable blocks, both depthwise-separable and blueprint-separable convolutions factorize a standard convolution into a pointwise ($1\times1$) and a depthwise step, substantially reducing multiply--accumulate operations. They differ in the order of factorization: depthwise-separable convolution applies the depthwise step before the pointwise one and is more efficient when the channel dimension expands, whereas blueprint-separable convolution applies the pointwise step first and is more efficient when the channel dimension is preserved or reduced. We therefore use depthwise-separable convolutions~\cite{depthwiseconv} (Fig.~\ref{fig:litecass_network}~(d)) at channel-expanding layers and blueprint-separable convolutions~\cite{blueprintconv} (Fig.~\ref{fig:litecass_network}~(c)) at channel-preserving or channel-reducing layers, so that each separable block attains the minimal computational cost. This hybrid design provides a lightweight mask estimator while maintaining sufficient capacity for speech, music, and SFX separation.

\subsection{Training objective}

LiteCASS is trained with a multi-task waveform-domain $\ell_1$ objective. Let $\mathbf{y}_i$ and $\hat{\mathbf{y}}_i$ denote the target and model output of source $i$, and let $\hat{\mathbf{y}}_{\mathrm{ns}}$ denote the intermediate non-speech output. The loss is defined as
\begin{equation}
\mathcal{L} =
\sum_{i \in \{\mathrm{sp},\mathrm{mu},\mathrm{sx}\}}
\|\hat{\mathbf{y}}_i-{\mathbf{y}}_i\|_1
+
\|\hat{\mathbf{y}}_{\mathrm{ns}}-({\mathbf{y}}_{\mathrm{mu}}+{\mathbf{y}}_{\mathrm{sx}})\|_1 ,
\label{eq:loss}
\end{equation}
where $\|\cdot\|_1$ denotes the mean absolute error over channels and time. The first three terms are the separation losses for speech, music, and SFX, while the last term regularizes the intermediate non-speech estimate toward the reference residual $\mathbf{y}_{\mathrm{mu}}+\mathbf{y}_{\mathrm{sx}}$.


  \begin{table*}[!t]
    \caption{Source separation performance on DnR\,v3-stereo.}
    \label{tab:separation_results}
    \centering
    \footnotesize
    \setlength{\tabcolsep}{3.2pt}
    \begin{tabular}{@{}lccccccccccccccc@{}}
      \toprule
      \multirow{2}{*}{\textbf{Model}}
      & \multicolumn{5}{c}{\textbf{Speech}}
      & \multicolumn{3}{c}{\textbf{Music}}
      & \multicolumn{3}{c}{\textbf{SFX}}
      & \multicolumn{3}{c}{\textbf{Averaged}} \\
      \cmidrule(lr){2-6}
      \cmidrule(lr){7-9}
      \cmidrule(lr){10-12}
      \cmidrule(lr){13-15}
      & \textbf{SDR} & \textbf{SI-SDR} & \textbf{SI-SDRi} & \textbf{PESQ} & \textbf{STOI}
      & \textbf{SDR} & \textbf{SI-SDR} & \textbf{SI-SDRi}
      & \textbf{SDR} & \textbf{SI-SDR} & \textbf{SI-SDRi}
      & \textbf{SDR} & \textbf{SI-SDR} & \textbf{SI-SDRi} \\
      \midrule
      MRX~\cite{mrx}
        & - & 11.35 & 13.57 & 1.96 & 0.834
        & - & 4.30 & 12.79
        & - & 4.49 & 13.74
        & - & 6.71 & 13.37 \\
      BandIt~\cite{bandit}
        & \textbf{14.50} & \textbf{13.91} & \textbf{16.13} & 2.19 & 0.869
        & \textbf{9.47} & 7.26 & 15.75
        & 8.74 & 7.11 & 16.37
        & \textbf{10.90} & 9.43 & 16.08 \\
      AV-CASS~\cite{avcass}
        & 12.40 & 12.51 & 13.45 & \textbf{2.41} & \textbf{0.900}
        & 8.43 & 7.19 & \textbf{16.11}
        & 8.86 & 7.95 & 16.96
        & 9.90 & 9.22 & 15.51 \\
      LiteCASS-K16
        & 13.02 & 12.33 & 14.55 & 1.94 & 0.845
        & 8.80 & 7.19 & 15.68
        & 9.05 & 7.57 & 16.83
        & 10.29 & 9.03 & 15.69 \\
      LiteCASS-K8
        & 13.54 & 12.93 & 15.15 & 2.11 & 0.865
        & 9.15 & \textbf{7.58} & 16.07
        & \textbf{9.38} & \textbf{8.00} & \textbf{17.26}
        & 10.69 & \textbf{9.50} & \textbf{16.16} \\
      \bottomrule
    \end{tabular}
  \end{table*}

\section{Experimental Procedures}
\label{sec:exp}

\subsection{Dataset}
\label{ssec:dataset}
To evaluate the proposed lightweight CASS model, we use Divide and Remaster v3 (DnR\,v3)~\cite{dnrv3}, which provides time-aligned speech, music, and SFX stems with additive mixtures. The dataset contains 6000 training clips, 600 validation clips, and 1200 test clips, each approximately 60\,s long.

Since cinematic audio is typically delivered in stereo or higher-channel formats, spatial cues such as panning and inter-channel delay should be preserved after separation. However, DnR\,v3 is distributed in mono. We therefore construct DnR\,v3-stereo by independently spatializing each mono stem with linear operations, including constant-power panning, Haas delay, and mid/side widening. The stereo mixture is then generated by summing the three spatialized stems, ensuring exact sum consistency. Speech is placed near the center, music uses wide-stereo patterns, and SFX is distributed across frontal and lateral positions. All audio is resampled to 48\,kHz and jointly peak-normalized to avoid clipping.

\subsection{Experiment configurations}
\label{ssec:config}

In our model, the STFT window length and hop size are $42.7$\,ms and $10.7$\,ms, respectively. The DFT length is 2048, and 1025 frequency bins are used for each frame. The subband factor is set to $K=8$ unless otherwise specified. Each Subband U-Net uses the channel configuration $[KC,32,64,64,64,64]$ and sigmoid mask activation.

We train on 512-frame segments for 200 epochs with a batch size of 64. The first 2 epochs are used for learning-rate warm-up. Adam~\cite{adam} is used as the optimizer. The peak learning rate is $1\times10^{-4}$, followed by cosine annealing to $1\times10^{-6}$. The loss weights are set to $\lambda_{\mathrm{sp}}=0.3$, $\lambda_{\mathrm{mu}}=1.0$, and $\lambda_{\mathrm{sx}}=1.0$. All models are trained on an NVIDIA A800 GPU.

\subsection{Baselines and Evaluation Metrics}
\label{sec:exp_baselines}

We compare LiteCASS with three CASS baselines, MRX~\cite{mrx}, BandIt~\cite{bandit}, and AV-CASS~\cite{avcass}, all trained from scratch on the DnR\,v3-stereo dataset of Section~\ref{ssec:dataset} under their original configurations. We report two variants, LiteCASS-K16 and LiteCASS-K8, with subband factors $K=16$ and $K=8$.

We assess separation with SDR, SI-SDR~\cite{convtasnet}, and SI-SDRi~\cite{sisdr} (higher is better), and additionally report PESQ~\cite{pesq} and STOI~\cite{stoi} for speech. To quantify stereo cue preservation, we adopt three inter-channel fidelity metrics computed as the deviations between estimated and reference sources: inter-channel level difference (ILD) error, inter-channel time difference (ITD) error~\cite{itd-ild}, and inter-channel coherence (ICC) error (lower is better). We also report parameters (Params), multiply--accumulate operations (MACs), and real-time factor (RTF), measured for separating a 10\,s, 48\,kHz stereo clip into three sources on an Intel Xeon Platinum 8358P CPU @ 2.60\,GHz.

\section{Results and Discussion}
\label{sec:results}
\subsection{Separation Performance}
\label{sec:exp_separation}

As shown in Table~\ref{tab:separation_results}, LiteCASS-K8 attains the best averaged SI-SDR ($9.50$\,dB) and SI-SDRi ($16.16$\,dB) and the best SFX scores among all systems, while using only $1.06$\,M parameters. BandIt and AV-CASS lead on speech reconstruction and perceptual quality, respectively, but LiteCASS remains competitive without any visual input. The smaller subband factor $K=8$ consistently outperforms $K=16$, confirming that $K$ trades frequency resolution against computation. We attribute this parameter--accuracy balance to three designs: the subband rearrangement shortens the convolutional frequency axis without discarding spectral information, so a shallow U-Net still covers the full band; the end-to-end hierarchy lets each U-Net specialize on speech or on the residual music--SFX mixture, rather than diluting capacity across all three stems; and the separable convolutions are confined to the two highest-resolution blocks that dominate computation, removing redundant computation rather than modeling power.

\subsection{Stereo Spatial Fidelity}
\label{sec:exp_spatial}

Table~\ref{tab:spatial_results} reports stereo spatial fidelity averaged over the three stems. We report inter-channel level difference (ILD) errors in dB, inter-channel time difference (ITD) errors in $\mu$s, and unitless inter-channel coherence (ICC) errors; lower values indicate better spatial cue preservation. LiteCASS-K8 yields lower ILD, ITD, and ICC errors than both audio-only baselines, MRX and BandIt, and is second only to AV-CASS, which exploits an additional visual modality. These results suggest that applying real-valued masks to the original complex spectrogram preserves the dominant inter-channel cues well. Because LiteCASS retains the mixture phase and does not explicitly reconstruct phase or spatial parameters, it maintains spatial fidelity while keeping the model lightweight.

\begin{table}[!t]
  \caption{Averaged stereo spatial fidelity on DnR\,v3-stereo
  (mean over speech, music, and SFX; lower is better).}
  \label{tab:spatial_results}
  \centering
  \footnotesize
  \setlength{\tabcolsep}{4.5pt}
  \begin{tabular}{@{}lccc@{}}
    \toprule
    \textbf{Model}
    & \makecell{\textbf{ILD err.}\\\textbf{(dB)}}
    & \makecell{\textbf{ITD err.}\\\textbf{($\mu$s)}}
    & \makecell{\textbf{ICC err.}\\\textbf{(unitless)}} \\
    \midrule
    MRX~\cite{mrx}        & 0.954 & 213.3 & 0.090 \\
    BandIt~\cite{bandit}  & 1.142 & 162.2 & 0.066 \\
    AV-CASS~\cite{avcass} & \textbf{0.335} & \textbf{118.9} & \textbf{0.028} \\
    LiteCASS-K16          & 0.409 & 145.1 & 0.054 \\
    LiteCASS-K8           & 0.401 & 139.3 & 0.053 \\
    \bottomrule
  \end{tabular}
\end{table}



\subsection{Model Complexity and Efficiency}
\label{sec:exp_complexity}
\begin{table}[!t]
  \caption{Model complexity and computational efficiency. ``--'' denotes CPU inference that could not be measured.}
  \label{tab:complexity}
  \centering
  \footnotesize
  \setlength{\tabcolsep}{4.0pt}
  \begin{tabular}{@{}lcccc@{}}
    \toprule
    \textbf{Model} & \textbf{Params(M)} & \textbf{MACs(G/s)} & \textbf{RTF$_\text{CPU}$} & \textbf{RTF$_\text{GPU}$} \\
    \midrule
    MRX~\cite{mrx}                & 36.30 & 9.05    & 0.094 & 0.019 \\
    BandIt~\cite{bandit}          & 37.02 & 126.18  & 1.339 & 0.025 \\
    AV-CASS~\cite{avcass}         & 30.16 & 1948.2  & --    & 0.352 \\
    LiteCASS-K16                  & \textbf{1.06}  & \textbf{0.38}    & \textbf{0.012} & \textbf{0.0011} \\
    LiteCASS-K8                   & \textbf{1.06}  & 0.72    & 0.014 & \textbf{0.0011} \\
    \bottomrule
  \end{tabular}
\end{table}

Table~\ref{tab:complexity} shows that LiteCASS is about $28$--$35\times$ smaller than the baselines and runs at an RTF of $0.012$ on a single CPU core, far below the real-time threshold, while BandIt and AV-CASS are far from real time on CPU. Together with Table~\ref{tab:separation_results}, this places LiteCASS on a favorable accuracy--efficiency frontier for real-time and embedded CASS.

\section{Conclusion}
\label{sec:conclusion}
We presented LiteCASS, a lightweight end-to-end network for real-time stereo CASS. It combines deterministic subband rearrangement with two compact U-Nets for hierarchical speech, music, and SFX separation. On a sum-consistent stereo extension of DnR\,v3, LiteCASS uses only $1.06$\,M parameters while achieving the best averaged SI-SDR and SFX separation among compared systems, with real-time CPU inference and competitive spatial fidelity.

\begingroup
\small
\renewcommand{\baselinestretch}{0.96}\selectfont
\bibliographystyle{IEEEbib}
\bibliography{mybib}

@string{atraslp   = "IEEE/ACM Trans. Audio, Speech, Lang. Process."}

@string{eusipco  = "Proc. Euro. Signal Process. Conf. (EUSIPCO)"}

@string{icassp   = "Proc. IEEE Int. Conf. Acoust., Speech, Signal Process."}

@string{cvpr = "Proc. IEEE Conf. Comput. Vis. Pattern Recognit."}

@article{bandit,
  title={A generalized bandsplit neural network for cinematic audio source separation},
  author={Watcharasupat, Karn N and Wu, Chih-Wei and Ding, Yiwei and Orife, Iroro and Hipple, Aaron J and Williams, Phillip A and Kramer, Scott and Lerch, Alexander and Wolcott, William},
  journal={IEEE Open J. Signal Process.},
  volume={5},
  pages={73--81},
  year={2023},
  publisher={IEEE}
}

@inproceedings{mrx,
  title={The cocktail fork problem: Three-stem audio separation for real-world soundtracks},
  author={Petermann, Darius and Wichern, Gordon and Wang, Zhong-Qiu and Le Roux, Jonathan},
  booktitle=icassp,
  pages={526--530},
  year={2022},
  organization={IEEE}
}

@inproceedings{dialogueenhancement,
  title={Dialogue enhancement of stereo sound},
  author={Geiger, J{\"u}rgen T and Grosche, Peter and Parodi, Yesenia Lacouture},
  booktitle=eusipco,
  pages={869--873},
  year={2015},
  organization={IEEE}
}

@article{loudnessnormalisation,
  title={Loudness normalisation and permitted maximum level of audio signals},
  author={EBU-Recommendation, R},
  journal={Eur. Broadcast. Union},
  year={2011}
}

@article{bsrnn,
  title={Music source separation with band-split RNN},
  author={Luo, Yi and Yu, Jianwei},
  journal=atraslp,
  volume={31},
  pages={1893--1901},
  year={2023},
  publisher={IEEE}
}

@inproceedings{htdemucs,
  title={Hybrid transformers for music source separation},
  author={Rouard, Simon and Massa, Francisco and D{\'e}fossez, Alexandre},
  booktitle=icassp,
  pages={1--5},
  year={2023},
  organization={IEEE}
}

@article{convtasnet,
  title={Conv-tasnet: Surpassing ideal time--frequency magnitude masking for speech separation},
  author={Luo, Yi and Mesgarani, Nima},
  journal=atraslp,
  volume={27},
  number={8},
  pages={1256--1266},
  year={2019},
  publisher={IEEE}
}

@inproceedings{ultralitess,
  title={Ultra-lightweight speech separation via group communication},
  author={Luo, Yi and Han, Cong and Mesgarani, Nima},
  booktitle=icassp,
  pages={16--20},
  year={2021},
  organization={IEEE}
}

@article{demucs,
  title={Music source separation in the waveform domain},
  author={D{\'e}fossez, Alexandre and Usunier, Nicolas and Bottou, L{\'e}on and Bach, Francis},
  journal={arXiv preprint arXiv:1911.13254},
  year={2019}
}

@article{tacklingcocktail,
  title={Tackling the cocktail fork problem for separation and transcription of real-world soundtracks},
  author={Petermann, Darius and Wichern, Gordon and Subramanian, Aswin Shanmugam and Wang, Zhong-Qiu and Le Roux, Jonathan},
  journal=atraslp,
  volume={31},
  pages={2592--2605},
  year={2023},
  publisher={IEEE}
}

@article{subtitle,
  title={Research and development of a subtitle management system using artificial intelligence.},
  author={Striuk, Andrii M and Hordiienko, Vladyslav V},
  journal={CS\&SE@ SW},
  pages={415--427},
  year={2024}
}

@inproceedings{fspen,
  title={Fspen: An ultra-lightweight network for real-time speech enhancement},
  author={Yang, Lei and Liu, Wei and Meng, Ruijie and Lee, Gunwoo and Baek, Soonho and Moon, Han-Gil},
  booktitle=icassp,
  pages={10671--10675},
  year={2024},
  organization={IEEE}
}

@article{ulunas,
  title={Ul-unas: Ultra-lightweight u-nets for real-time speech enhancement via network architecture search},
  author={Rong, Xiaobin and Yang, Leyan and Wang, Dahan and Hu, Yuxiang and Zhu, Changbao and Chen, Kai and Lu, Jing},
  journal=atraslp,
  year={2026},
  publisher={IEEE}
}

@article{adam,
  title={Adam: A method for stochastic optimization},
  author={Kingma, Diederik P and Ba, Jimmy},
  journal={arXiv preprint arXiv:1412.6980},
  year={2014}
}

@inproceedings{dnrv3,
  title={Remastering divide and remaster: A cinematic audio source separation dataset with multilingual support},
  author={Watcharasupat, Karn N and Wu, Chih-Wei and Orife, Iroro},
  booktitle={Proc. IEEE 5th Int. Symp. Internet of Sounds (IS2)},
  pages={1--10},
  year={2024},
  organization={IEEE}
}

@inproceedings{avcass,
  title={Cinematic audio source separation using visual cues},
  author={Zhang, Kang and Lee, Suyeon and Senocak, Arda and Chung, Joon Son},
  booktitle=cvpr,
  pages={37874--37884},
  year={2026}
}

@article{hdemucs,
  title={Hybrid spectrogram and waveform source separation},
  author={D{\'e}fossez, Alexandre},
  journal={arXiv preprint arXiv:2111.03600},
  year={2021}
}

@inproceedings{pesq,
  title={Perceptual evaluation of speech quality (PESQ)-a new method for speech quality assessment of telephone networks and codecs},
  author={Rix, Antony W and Beerends, John G and Hollier, Michael P and Hekstra, Andries P},
  booktitle=icassp,
  volume={2},
  pages={749--752},
  year={2001},
  organization={IEEE}
}

@inproceedings{stoi,
  title={A short-time objective intelligibility measure for time-frequency weighted noisy speech},
  author={Taal, Cees H and Hendriks, Richard C and Heusdens, Richard and Jensen, Jesper},
  booktitle=icassp,
  pages={4214--4217},
  year={2010},
  organization={IEEE}
}

@inproceedings{sisdr,
  title={SDR--half-baked or well done?},
  author={Le Roux, Jonathan and Wisdom, Scott and Erdogan, Hakan and Hershey, John R},
  booktitle=icassp,
  pages={626--630},
  year={2019},
  organization={IEEE}
}

@InProceedings{casstask,
  author =	 {Petermann, Darius and Wichern, Gordon and Wang, Zhong-Qiu and {Le Roux}, Jonathan},
  title =	 {The Cocktail Fork Problem: Three-Stem Audio Separation for Real-World Soundtracks},
  booktitle =icassp,
  year =	 2022,
  month =	 may
}

@article{itd-ild,
  title={XII. On our perception of sound direction},
  author={Rayleigh, Lord},
  journal={The London, Edinburgh, and Dublin Philosophical Magazine and Journal of Science},
  volume={13},
  number={74},
  pages={214--232},
  year={1907},
  publisher={Taylor \& Francis}
}

@INPROCEEDINGS{crossnetunmix,
  author={Sawata, Ryosuke and Uhlich, Stefan and Takahashi, Shusuke and Mitsufuji, Yuki},
  booktitle=icassp, 
  title={All For One And One For All: Improving Music Separation By Bridging Networks}, 
  year={2021},
  volume={},
  number={},
  pages={51-55},
  doi={10.1109/ICASSP39728.2021.9414044}}

@article{gru,
  title={Empirical evaluation of gated recurrent neural networks on sequence modeling},
  author={Chung, Junyoung and Gulcehre, Caglar and Cho, KyungHyun and Bengio, Yoshua},
  journal={arXiv preprint arXiv:1412.3555},
  year={2014}
}

@inproceedings{tcn,
  title={Temporal convolutional networks for action segmentation and detection},
  author={Lea, Colin and Flynn, Michael D and Vidal, Rene and Reiter, Austin and Hager, Gregory D},
  booktitle={proceedings of the IEEE Conference on Computer Vision and Pattern Recognition},
  pages={156--165},
  year={2017}
}

@inproceedings{blueprintconv,
  title={Blueprint separable residual network for efficient image super-resolution},
  author={Li, Zheyuan and Liu, Yingqi and Chen, Xiangyu and Cai, Haoming and Gu, Jinjin and Qiao, Yu and Dong, Chao},
  booktitle=cvpr,
  pages={833--843},
  year={2022}
}

@article{depthwiseconv,
  title={Depthwise separable convolutions with deep residual convolutions},
  author={Hasan, Md Arid and Dey, Krishno},
  journal={arXiv preprint arXiv:2411.07544},
  year={2024}
}
\endgroup

\end{document}